\documentclass[conference]{IEEEtran}
\IEEEoverridecommandlockouts

\usepackage{cite}
\usepackage{amsmath,amssymb,amsfonts}
\usepackage{xspace}
\usepackage{graphicx}
\usepackage{listings}
\usepackage{xcolor}
\usepackage{multirow}
\usepackage{float}
\usepackage[utf8]{inputenc}
\usepackage[most]{tcolorbox}
\usepackage{enumitem}
\usepackage{booktabs}
\usepackage{url}

\usepackage[htt]{hyphenat}

\definecolor{findingbg}{RGB}{255,248,210}

\newtcolorbox{findingbox}{
    enhanced,
    before skip=3mm,
    after skip=3mm,
    colback=findingbg,
    colframe=findingbg,
    boxrule=0pt,
    sharp corners,
    left=4mm,
    right=4mm,
    top=2mm,
    bottom=2mm
}

\tcbset{
  promptbox/.style={
    enhanced,
    colback=gray!5,
    colframe=gray!50,
    boxrule=0.5pt,
    arc=2mm,
    outer arc=2mm,
    fonttitle=\bfseries,
    coltitle=black,
    title={#1},
    sharp corners=south,
    breakable,
    listing only,
    listing options={basicstyle=\ttfamily\footnotesize, breaklines=true, breakindent=0pt}
  }
}

\newcommand{\up}[1]{\rlap{\;\scriptsize\textcolor{green!60!black}{($\uparrow$#1)}}}
\newcommand{\down}[1]{\rlap{\;\scriptsize\textcolor{red!70!black}{($\downarrow$#1)}}}

\definecolor{keycolor}{rgb}{0.0, 0.0, 0.8}
\definecolor{valuecolor}{rgb}{0.8, 0.1, 0.1}

\def\BibTeX{{\rm B\kern-.05em{\sc i\kern-.025em b}\kern-.08em
    T\kern-.1667em\lower.7ex\hbox{E}\kern-.125emX}}

\newcommand{\approach}{{ACQUIRE}\xspace}
\newcommand{\questionertag}{Questioner\xspace}
\newcommand{\explorertag}{Answerer\xspace}
\newcommand{\resolvertag}{Resolver\xspace}

\begin{document}

\title{Know Before Fix: QA-Driven Repository Knowledge Acquisition for Software Issue Resolution}

\author{
\IEEEauthorblockN{
Haotian Lin\textsuperscript{1},
Silin Chen\textsuperscript{1},
Xiaodong Gu\textsuperscript{1*},
Yuling Shi\textsuperscript{1},\\
Chengxi Pan\textsuperscript{2},
Jiaqi Ge\textsuperscript{3},
Mengfan Li\textsuperscript{1},\\
Jianghong Huang\textsuperscript{1},
MENGCHIEH CHUANG\textsuperscript{1},
Beijun Shen\textsuperscript{1}, and
Haibing Guan\textsuperscript{1}}
\IEEEauthorblockA{\textsuperscript{1}Shanghai Jiao Tong University, Shanghai, China}
\IEEEauthorblockA{\textsuperscript{2}University of Pittsburgh, Pittsburgh, PA, USA}
\IEEEauthorblockA{\textsuperscript{3}Guangdong Technion--Israel Institute of Technology, Shantou, China}
\IEEEauthorblockA{
\{hyperlynnx, xiaodong.gu, yuling.shi, lmf2951510526, jh.huang, Zhuangmengjie, bjshen, hbguan\}@sjtu.edu.cn,\\
csl2457029646@163.com, chp252@pitt.edu, ge24876@gtiit.edu.cn}
\thanks{* Xiaodong Gu is the corresponding author.}
}

\maketitle

\begin{abstract}
LLM-based coding agents have significantly advanced automated software issue resolution, yet they remain highly prone to factual errors caused by insufficient repository understanding. Recent methods attempt to mitigate this limitation through pre-repair repository exploration; however, their fix-driven strategies explore repositories without identifying the agent's knowledge gaps, often yielding imprecise context that fails to bridge the underlying understanding deficit. In this paper, we propose \approach, a QA-driven framework for software issue resolution. Mirroring how experienced developers first comprehend unfamiliar code before attempting a fix, \approach explicitly acquires repository knowledge prior to repair. The framework decouples knowledge acquisition from patch generation through two stages: in the first stage, a \emph{\questionertag} and an \emph{\explorertag} collaborate to acquire structured repository knowledge, where the \emph{\questionertag} poses targeted questions and the \emph{\explorertag} produces evidence-grounded answers through autonomous exploration; in the second stage, the \emph{\resolvertag} leverages the resulting QA knowledge to generate informed patches. By transforming implicit knowledge gaps into explicit, factually reliable understanding, \approach accelerates knowledge-intensive repair stages and enables more accurate resolution. Experiments on SWE-bench Verified demonstrate that \approach consistently outperforms representative pre-repair methods, raising Pass@1 by up to 4.4 percentage points with modest additional cost and time.
\end{abstract}

\begin{IEEEkeywords}
software issue resolution, repository knowledge acquisition, coding agents, large language models
\end{IEEEkeywords}

\begin{figure*}[t]
  \centering
  \includegraphics[width=0.93\textwidth]{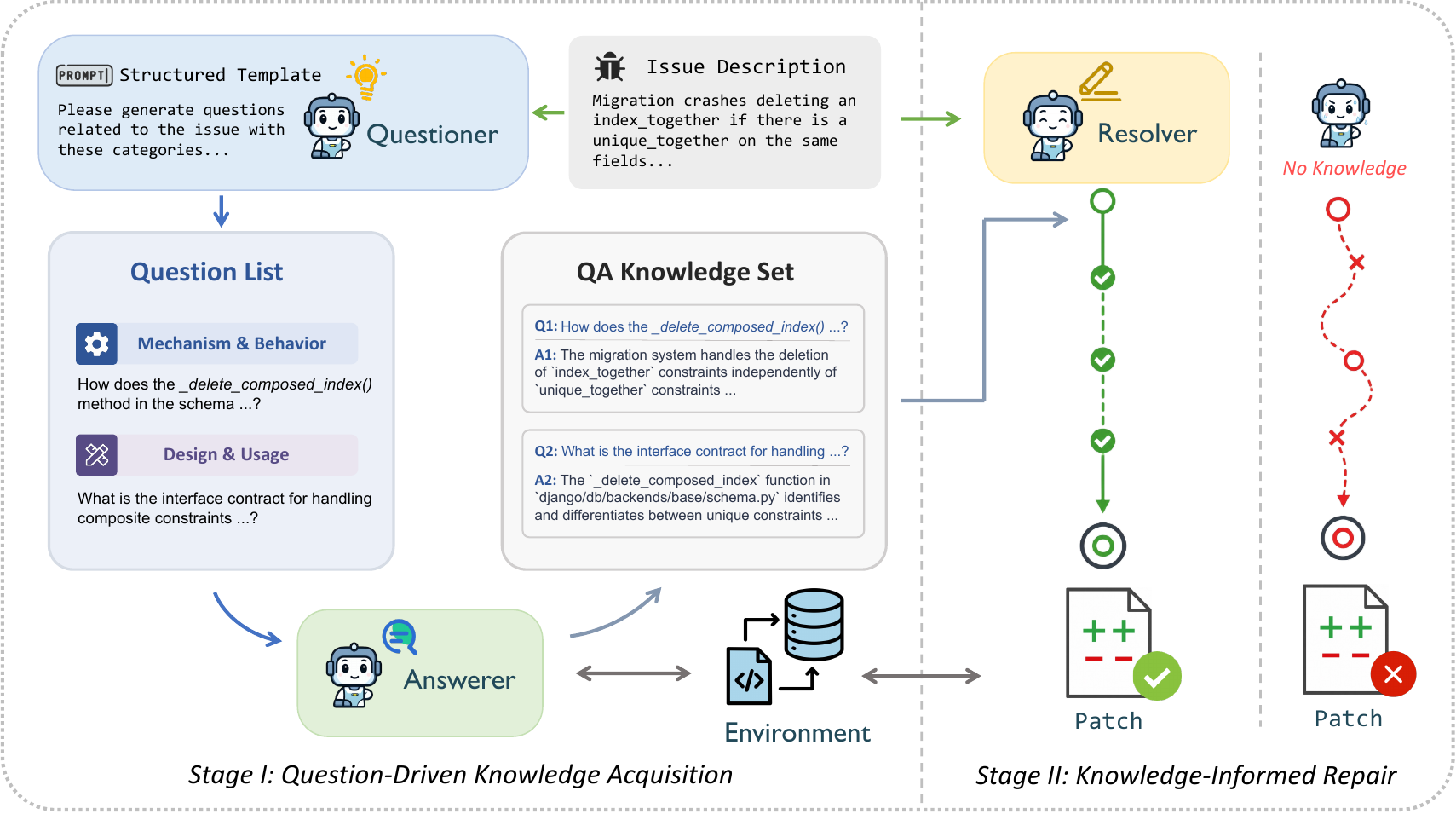}
  \caption{Overview of \approach.}
  \label{fig:framework}
\end{figure*}

\section{Introduction}
Coding agents empowered by large language models (LLMs) have driven significant advances in automated software issue resolution~\cite{li2025swe,wang2026swe,locagent}. Powered by techniques such as chain-of-thought reasoning, multi-step planning, and tool-augmented interaction, these agents can navigate complex codebases, identify relevant code locations, and generate candidate patches with remarkable fluency~\cite{yang2024sweagenta,agentless,autocoderover}.

Despite these rapid advancements, a critical class of failures remains largely unresolved, even for frontier-scale models: \emph{factual errors} stemming from \emph{insufficient repository understanding}.
These failures are not due to limitations in the model's reasoning capacity; rather, they arise because the issue description alone does not supply the repository-internal knowledge, such as cross-module dependencies, implicit API contracts, and data-flow details, required to formulate a correct fix. Consequently, agents often fall back on shallow, keyword-based localization, fail to trace fault origins across module boundaries, and inadvertently violate implicit API contracts~\cite{liu2025empiricalfailures,bouzenia2025understanding,autocoderover,he2026sweadept}. Furthermore, these knowledge-deficient attempts are disproportionately costly, consuming over four times the token cost and nearly twice the execution steps compared to successful resolutions~\cite{fan2025sweeffi,bouzenia2025understanding}.

To bridge this gap, recent methods have attempted to explore the repository prior to issue resolution~\cite{locagent,CoSIL,lingmaagent,li2025swe}. For instance, LingmaAgent links issue keywords to code entities and constructs structural summaries to provide broader codebase context~\cite{lingmaagent}. While effective for straightforward bugs, these techniques are fundamentally driven by issue keywords rather than by deep understanding of the repository, and the resulting context is frequently incomplete or imprecise, ultimately compounding the agent's misunderstanding of the codebase~\cite{he2026sweadept,liu2025empiricalfailures}.

We propose \textbf{\approach} (\textbf{A}gent \textbf{C}ollaboration for \textbf{Q}uestion-Answer-driven \textbf{I}ssue \textbf{RE}solution), a framework that explicitly decouples repository knowledge acquisition from patch generation. Mirroring how experienced developers first build understanding of an unfamiliar codebase before attempting a fix, \approach explicitly identifies what repository knowledge the agent lacks and \emph{proactively} acquires it before repair begins. The framework orchestrates three specialized agents across two stages. In the knowledge acquisition stage, a \questionertag and an \explorertag collaborate to acquire structured repository knowledge: the \questionertag decomposes the issue into targeted questions across complementary knowledge dimensions, and the \explorertag resolves each question through autonomous repository exploration, producing evidence-grounded QA pairs. In the repair stage, the \resolvertag leverages this structured QA knowledge to generate informed patches. This design transforms implicit knowledge gaps into explicit, structured understanding before editing begins.

Experiments on SWE-bench Verified~\cite{openai2024swebenchverified}, a curated benchmark of 500 real-world GitHub issues, demonstrate that \approach raises Pass@1 by up to 4.4 percentage points over the base repair agent, consistently outperforming all representative pre-repair methods across two backbone LLMs at modest additional cost and time.

Our contributions are threefold:
\begin{itemize}[leftmargin=2em, itemsep=4pt, topsep=2pt]
    \item We propose \approach{}, a QA-driven framework for software issue resolution. To our knowledge, this is the first work to integrate repository-level QA into issue resolution.
    \item \approach decouples knowledge acquisition from patch generation into two explicit stages, and introduces a category-guided question taxonomy that enables targeted acquisition of the repository knowledge needed for repair.
    \item We provide an in-depth analysis showing that the acquired knowledge is factually reliable, reduces repair trajectory length, and steers the agent toward causally relevant code regions on previously failed instances, revealing how pre-repair knowledge improves resolution.
\end{itemize}

\section{Motivation}
\label{sec:motivation}

Recent empirical studies~\cite{liu2025empiricalfailures,bouzenia2025understanding,autocoderover,he2026sweadept} have consistently identified \textbf{insufficient repository knowledge} as a dominant cause of SE agent failure, manifesting as shallow keyword-based localization, failure to trace faults across module boundaries, violation of implicit API contracts, and missing knowledge of external protocols~\cite{liu2025empiricalfailures,autocoderover,he2026sweadept}. Existing pre-repair methods attempt to mitigate this gap but remain fundamentally \emph{fix-oriented}: whether producing ranked suspicious locations~\cite{locagent,CoSIL} or enriching context with structural summaries~\cite{lingmaagent,li2025swe}, they rely on the agent's own ability to convert shallow pointers into correct patches, and neither family explicitly identifies \emph{what repository knowledge the agent still lacks} before repair begins~\cite{he2026sweadept,liu2025empiricalfailures}.
This contrasts with how experienced developers approach unfamiliar bugs: rather than immediately searching for a fix, they first ask diagnostic questions about the code, such as how a component behaves, what contracts an API enforces, or where relevant modules reside, building the necessary understanding before committing to any repair strategy~\cite{vessey1985expertise,gugerty1986debugging}.

Inspired by recent work on repository-level question answering~\cite{sweqa}, we pose a different question: if an agent could decompose an issue into targeted, \emph{easy-to-answer} repository questions and obtain accurate answers \emph{before} any repair attempt, could such high-quality knowledge more effectively unlock its repair potential? To test this hypothesis, we design an \emph{oracle QA} experiment, adopting the oracle probing setup of~\cite{ambigswe}.
Concretely, we construct an oracle \questionertag that takes the issue description together with the golden test patch as input and generates a single question most relevant to resolving the issue. An oracle \explorertag then receives only the files involved in the golden patch and uses them as grounding context to answer the question. We inject the resulting oracle QA pair into Mini-SWE-Agent~\cite{mini_swe_agent} and evaluate on the 116 instances from SWE-bench Lite~\cite{jimenez2023swe} that Mini-SWE-Agent fails to resolve under DeepSeek-V3.2~\cite{deepseek2025v32}.
With oracle QA injection, 26 of these 116 previously failed instances are successfully resolved, indicating that targeted pre-repair knowledge can bridge the gap between the agent's reasoning capacity and the repository understanding required for correct fixes. Note that this oracle setting is intentionally constrained, as only one question is asked and the oracle \explorertag can only access the gold files rather than the full repository, so the 26/116 recovery rate reflects this restricted setup rather than an inherent ceiling of the QA paradigm.

While the oracle experiment confirms the promise of question-driven knowledge injection, it relies on privileged information, namely the golden test patch and the golden file set, which is unavailable in real-world settings. To bridge this gap toward a fully automated pipeline, we address each oracle dependency in turn. For the \questionertag, five computer science graduate students manually analyzed the 116 oracle questions from the above experiment and, cross-referencing with the failure taxonomies established in the aforementioned studies~\cite{liu2025empiricalfailures,bouzenia2025understanding,autocoderover,he2026sweadept}, identified four recurring question patterns that cover distinct dimensions of the repository knowledge needed for repair:
(1)~\emph{Mechanism \& Behavior} (70.7\%), targeting how internal logic flows and data transformations actually work;
(2)~\emph{Design \& Usage} (18.1\%), targeting what API contracts and design conventions govern valid edits;
(3)~\emph{Locating \& Structure} (7.8\%), targeting where relevant code resides and how modules are organized;
and (4)~\emph{Ecosystem \& Standards} (3.4\%), targeting what external library behaviors or protocol constraints apply.
Together, they serve as the structured question template, enabling the \questionertag to generate targeted questions without any oracle information. For the \explorertag, we replace the privileged file set with an autonomous agent that explores the repository, discovering relevant files and gathering context on its own. The detailed design of both components is presented in Section~\ref{sec:methodology}.

\section{Methodology}
\label{sec:methodology}
\subsection{Overview}
We propose \approach, a two-stage framework that mirrors how experienced developers approach an unfamiliar codebase: first acquiring the necessary repository knowledge, then performing an informed repair. As illustrated in Figure~\ref{fig:framework}, given an issue description $\mathcal{I}$ and an execution environment $\mathcal{E}$, the \questionertag first generates $N$ targeted questions, and $N$ independently instantiated \explorertag instances explore the repository in parallel to answer them, producing a knowledge set $\mathcal{K} = \{(q_1, a_1), \ldots, (q_N, a_N)\}$. The \resolvertag then leverages $\mathcal{I}$ and $\mathcal{K}$ to generate a candidate patch. The following subsections detail each stage.

\subsection{Stage~1: Question-Driven Knowledge Acquisition}
\label{sec:stage1}

Stage~1 acquires structured repository knowledge through $N$ question--answer interactions before any repair action begins.

\subsubsection{Question Generation}
To ensure the questions cover knowledge dimensions most relevant to repair, the \questionertag is guided by a structured prompt template derived from the knowledge-deficit patterns identified in Section~\ref{sec:motivation}. The template encodes the four category descriptions with concrete examples, a JSON output schema, and constraints enforcing non-redundancy and self-containedness (the full template is provided in the supplementary material). The four categories are defined as follows, where the first three target distinct dimensions of \emph{repository-internal} knowledge, while the fourth addresses \emph{external} knowledge that the repository depends on but does not contain:

\begin{itemize}[leftmargin=*, itemsep=4pt, topsep=2pt]
  \item \textbf{Mechanism \& Behavior.} Targets functional logic flows, state management, and data processing details, asking \emph{how a functionality actually works}, e.g., ``How does the authentication system validate user credentials in this repository?'' Such questions are critical for tracing the root cause of behavioral bugs.

  \item \textbf{Design \& Usage.} Targets API definitions, class hierarchies, error specifications, and design conventions, asking \emph{how interfaces are designed and what contracts they follow}, e.g., ``What is the interface contract for the database connection class?'' These questions help the downstream agent avoid violating existing design constraints.

  \item \textbf{Locating \& Structure.} Targets the codebase layout, module organization, and dependency relationships, asking \emph{where things are and how they relate}, e.g., ``Where is the login functionality implemented?'' These questions directly reduce the search space for repair.

  \item \textbf{Ecosystem \& Standards.} Targets external knowledge such as third-party library behaviors, protocol specifications, and language-level semantics, asking \emph{what external knowledge is needed}, e.g., ``What does the HTTP/1.1 specification require for chunked transfer encoding?'' These questions help the repair agent reason correctly about constraints beyond the codebase.
\end{itemize}

The four categories serve as a guiding taxonomy rather than a rigid per-question constraint. The \questionertag autonomously selects the most appropriate category for each question based on the issue context; no external scheduling or round-robin policy is imposed. Consequently, a single category may recur across questions when the most critical knowledge gaps fall under it.

\subsubsection{Evidence-Grounded Answering}
For each question, the \explorertag produces a grounded answer through autonomous repository exploration. The \explorertag shares the same scaffolded shell interaction environment as Mini-SWE-Agent but operates in read-only mode, ensuring that no repository state is modified during knowledge acquisition. Its prompt specifies three behavioral requirements: (1)~answers must reference concrete repository artifacts such as file paths, function names, and relevant code behaviors, rather than relying on speculation or parametric knowledge; (2)~the agent should actively submit its answer once sufficient evidence has been gathered; and (3)~if sufficient evidence cannot be found, the agent should explicitly acknowledge the gap rather than fabricate unsupported claims. In practice, all instances completed within the allocated budget without forced truncation. The factual reliability of the acquired answers is evaluated in Section~\ref{sec:qa_reliability}, and a characterization of the generated QA is provided in the supplementary material.

\subsubsection{Parallel Knowledge Acquisition}
\label{sec:parallel_acq}
Given the issue description $\mathcal{I}$, the \questionertag generates $N$ questions $\{q_1, \ldots, q_N\}$. All $N$ questions are then dispatched to $N$ independently instantiated \explorertag instances that explore $\mathcal{E}$ in parallel, each producing a grounded answer $a_i$ for its assigned question $q_i$.

Each \explorertag instance receives only $\mathcal{I}$ and its assigned question $q_i$ as input, without access to other questions or exploration traces. This isolation serves two purposes: it prevents cross-instance context from biasing exploration toward previously visited but potentially irrelevant code regions, keeping each answer focused and independently grounded; and it enables fully parallel execution, reducing the wall-clock latency of the knowledge acquisition stage to that of the single slowest instance rather than the sum of all instances. After all $N$ instances complete, the resulting pairs are assembled into the knowledge set $\mathcal{K} = \{(q_1, a_1), \ldots, (q_N, a_N)\}$ and passed to Stage~2.

\subsection{Stage~2: Knowledge-Informed Repair}
\label{sec:stage2}

In Stage~2, the \resolvertag performs issue resolution guided by the repository knowledge acquired in Stage~1. The \resolvertag receives the issue description $\mathcal{I}$ together with the knowledge set $\mathcal{K}$, and operates within $\mathcal{E}$ through an iterative loop of code navigation, editing, and test execution until a candidate patch is produced.

The $N$ QA pairs are serialized into a structured text block and prepended to the \resolvertag's messages before the repair instruction, so that repository understanding is fully established before the first repair action and remains stable throughout the trajectory. We adopt static pre-injection rather than dynamic interleaving (i.e., streaming QA content into the trajectory on demand) for two reasons: (1)~it preserves the modularity of the two-stage design by avoiding coupling between knowledge acquisition and the repair loop; and (2)~it ensures the complete knowledge context is available from the first repair step, preventing early-stage decisions from being made under partial information. Importantly, the injected QA serves as supplementary rather than prescriptive context, informing the \resolvertag's strategies without overriding its ability to independently verify and adjust based on direct repository observations. This minimal-intervention design ensures that any performance difference is attributable to the acquired knowledge rather than to changes in the repair mechanism itself. Trajectory analysis shows that the agent consults QA knowledge primarily in knowledge-intensive phases, using it to narrow file search during localization and to inform edit decisions during fixing, while relying more on executable feedback in reproduction and verification phases (Section~\ref{sec:stage2_analysis}).

\section{Experimental Setup}
\subsection{Research Questions}
We evaluate \approach by addressing the following research questions:

\textbf{RQ1 (Effectiveness):} How effective is \approach for issue resolution?

\textbf{RQ2 (Knowledge Quality \& Influence):} How effectively does the generated QA knowledge support repair behavior?

\textbf{RQ3 (Ablation):} How much do the question decomposition and the category-guided question generation each contribute to repair?
\textbf{RQ4 (Number of QA):} How does the number of QA pairs affect repair performance and cost?

\subsection{Datasets and Models}
We evaluated our method on SWE-bench Verified~\cite{openai2024swebenchverified}, a high-quality subset of SWE-bench~\cite{jimenez2023swe}, containing 500 real GitHub issues focused on fixing functional bugs in a controlled, isolated environment. For each instance, the model only receives a problem description in natural language and the corresponding code repository. The correctness of the generated patches is evaluated by running unit tests written by the developers, thus providing a consistent and rigorous evaluation of the performance of automated bug fixing.

To assess the generality of \approach across model families, we adopt two backbone LLMs that have been widely used to evaluate SE agents~\cite{guo2026eet,suri2026codescout, xia2025live,hayashi2025sage}, representing distinct development paradigms:

\begin{itemize}[leftmargin=*, itemsep=4pt, topsep=2pt]
  \item \textbf{DeepSeek-V3.2}~\cite{deepseek2025v32}: an open-source model built on a Mixture-of-Experts (MoE) architecture with 671B total parameters and approximately 37B activated per token. It incorporates sparse attention mechanisms and large-scale reinforcement learning post-training, achieving frontier-level performance in code understanding and generation tasks.
  \item \textbf{GPT-5-mini}~\cite{openai2025gpt5mini}: a proprietary, efficiency-oriented model from OpenAI, optimized for high throughput and low-latency deployment while maintaining strong general-purpose capabilities.
\end{itemize}
By pairing an open-source reasoning-specialized model with a closed-source efficiency-oriented model, our evaluation covers two representative points in the current LLM landscape.

\subsection{Evaluation Metrics}
We evaluate all methods using two primary metrics that jointly capture resolution effectiveness and economic cost:
\begin{itemize}[leftmargin=*, itemsep=2pt, topsep=2pt]
  \item \textbf{Pass@1}: the proportion of instances successfully resolved in a single attempt, serving as the primary effectiveness metric.
  \item \textbf{Average Cost}: the average monetary cost per instance over the entire pipeline, which includes both the pre-repair exploration stage and the repair stage.
  \item \textbf{Average Time}: the average end-to-end wall-clock time per instance, measured from the start of the pre-repair stage to the completion of patch generation.
\end{itemize}

\subsection{Baseline Methods}
We compare \approach with representative pre-repair exploration methods, alongside Mini-SWE-Agent which serves as the shared base repair agent.

\begin{itemize}[leftmargin=*, itemsep=4pt, topsep=2pt]
  \item \textbf{Mini-SWE-Agent}~\cite{mini_swe_agent}: a minimal agentic repair system derived from SWE-agent~\cite{yang2024sweagenta} that iteratively proposes shell actions, observes execution results, and updates its strategy until a patch is produced. It has been widely adopted as a standardized repair backbone in recent evaluations~\cite{guo2026eet, wang2026swe, fan2025sweeffi, he2026sweadept}; in our study, it serves both as a standalone baseline and as the shared base repair agent for all compared methods. Its minimal design makes it straightforward to attribute performance differences to the methods themselves.

  \item \textbf{LocAgent}~\cite{locagent}: constructs a heterogeneous repository graph encoding import, call, and inheritance relations, then performs LLM-guided multi-hop traversal to produce a ranked list of suspicious fault locations.

  \item \textbf{CoSIL}~\cite{CoSIL}: incrementally expands a local call graph during search, iteratively refining the search frontier and pruning context to yield compact, high-relevance code fragments as pre-repair evidence.

  \item \textbf{LingmaAgent}~\cite{lingmaagent}: builds a repository-level knowledge graph with a summarized global view, then uses Monte Carlo Tree Search (MCTS) to guide exploration, producing fine-grained fault localization signals and a repair-oriented repository summary.

  \item \textbf{SWE-Debate}~\cite{li2025swe}: introduces multi-agent debate into the repair pipeline, where multiple LLM agents independently propose repair hypotheses, iteratively challenge each other's reasoning, and converge on a consensus repair plan that guides downstream patch generation.
\end{itemize}

\subsection{Implementation Details}

\subsubsection{Baseline Method Configurations}
For \textbf{Mini-SWE-Agent}, we follow the official SWE-bench setup. The agent runs in an isolated Docker container, a per-step command timeout of 360 seconds, a maximum trajectory length of 250 steps, a per-instance cost cap of \$3.00, and temperature 0.0. In our implementation, all baselines and \approach adopt Mini-SWE-Agent as the base repair agent and share the same repair-stage settings. For all other baseline methods, we follow the default settings reported in their original papers.

\subsubsection{\approach Configurations}
The knowledge acquisition stage generates $N{=}2$ questions per issue, with \questionertag using temperature 0.7 to encourage diverse and complementary coverage across the four question categories. \explorertag uses the same scaffolded interaction environment as Mini-SWE-Agent with read-only execution, a per-step timeout of 120 seconds, a maximum trajectory length of 150 steps, a per-instance cost cap of \$2.00, and temperature 0.0. \resolvertag inherits the same scaffold, execution environment, and settings as the baseline Mini-SWE-Agent. Due to space limitations, the prompt templates are provided in the supplementary material.

\begin{table}[!t]
  \caption{Main results on SWE-bench Verified under GPT-5-mini and DeepSeek-V3.2.}
  \centering
  \small
  \resizebox{\columnwidth}{!}{%
  \begin{tabular}{llccc}
  \toprule
  \textbf{Method} & \textbf{Model} & \textbf{Pass@1 (\%)} & \textbf{Avg Cost (\$)} & \textbf{Avg Time (s)} \\
  \midrule
  \multirow{2}{*}{Mini-SWE-Agent} & GPT-5-mini    & 58.4 & 0.024 & 187 \\
                                   & DeepSeek-V3.2 & 66.4 & 0.055 & 815 \\
  \midrule
  \multirow{2}{*}{LocAgent}       & GPT-5-mini    & 57.8 \down{0.6} & 0.048 & 437 \\
                                   & DeepSeek-V3.2 & \underline{68.8} \up{2.4} & 0.060 & 1046 \\
  \midrule
  \multirow{2}{*}{CoSIL}          & GPT-5-mini    & 55.2 \down{3.2} & 0.035 & 224 \\
                                   & DeepSeek-V3.2 & 68.7 \up{2.3} & 0.059 & 750 \\
  \midrule
  \multirow{2}{*}{LingmaAgent}    & GPT-5-mini    & \underline{60.0} \up{1.6} & 0.309 & 823 \\
                                   & DeepSeek-V3.2 & 67.4 \up{1.0} & 0.155 & 1421 \\
  \midrule
  \multirow{2}{*}{SWE-Debate}     & GPT-5-mini    & 59.8 \up{1.4} & 0.738 & 1552 \\
                                   & DeepSeek-V3.2 & 68.2 \up{1.8} & 0.382 & 2517 \\
  \midrule
  \multirow{2}{*}{\textbf{\approach (Ours)}} & GPT-5-mini    & \textbf{62.2} \up{3.8} & 0.054 & 302 \\
                                      & DeepSeek-V3.2 & \textbf{70.8} \up{4.4} & 0.073 & 1042 \\
  \bottomrule
  \end{tabular}}
  \label{tab:main_results}
\end{table}

\section{Results}

\subsection{RQ1: Effectiveness of \approach}

Table~\ref{tab:main_results} reports the main results. Arrows indicate improvement or degradation relative to Mini-SWE-Agent, and underlines mark the best-performing baseline for each backbone model.

Among the baselines, LocAgent and CoSIL achieve moderate gains under DeepSeek-V3.2 but degrade under GPT-5-mini. This cross-model inconsistency highlights a limitation of localization-oriented approaches: they provide shallow code pointers (e.g., suspicious file lists or code fragments) and rely on the backbone model to infer \emph{why} these locations are relevant. When the model lacks this reasoning capacity, such signals can mislead the repair process. In contrast, LingmaAgent and SWE-Debate combine static analysis with LLM-driven repository traversal. LingmaAgent employs MCTS-based planning to produce repair-oriented summaries, while SWE-Debate leverages multi-agent debate to converge on a consensus repair plan. Both yield consistent improvements across models, confirming the value of richer pre-repair context. However, their extensive traversal substantially increases cost and time, limiting practical deployment.

\approach outperforms all baselines by a clear margin. First, it consistently achieves the largest Pass@1 gains across both backbone models (+3.8 on GPT-5-mini, +4.4 on DeepSeek-V3.2), demonstrating strong cross-model generalization; in contrast, localization-oriented baselines even degrade under GPT-5-mini.
Second, unlike localization baselines that supply shallow code pointers, the QA-driven paradigm captures richer knowledge dimensions such as behavioral context, design constraints, and structural relationships, as further characterized in the supplementary material. While LingmaAgent and SWE-Debate also produce richer context, they do so at 1.4--5$\times$ longer end-to-end time and 2--14$\times$ higher cost.
Third, \approach maintains competitive efficiency, with end-to-end time and cost comparable to the lightweight localization baselines. This favorable accuracy--efficiency trade-off stems from the QA-driven paradigm, which elicits targeted knowledge without exhaustive repository traversal.

\begin{findingbox}
\textbf{Finding 1:} \approach achieves the highest Pass@1 across both GPT-5-mini (+3.8) and DeepSeek-V3.2 (+4.4), with modest cost and time overhead. These results demonstrate that targeted QA-driven knowledge acquisition is more effective, more generalizable, and more efficient than existing pre-repair strategies.
\end{findingbox}

\subsection{RQ2: Quality and Influence of QA Knowledge}
\label{sec:rq2}

Having established that \approach consistently improves resolution rates, this section examines the reliability of the generated QA knowledge and its influence on downstream repair behavior. To control for model variation, all analyses in this section use DeepSeek-V3.2.

\subsubsection{QA Factual Reliability}
\label{sec:qa_reliability}

As shown in Table~\ref{tab:main_results}, comparing per-instance outcomes between Mini-SWE-Agent and \approach yields 44 Fail$\to$Pass recoveries and 22 Pass$\to$Fail regressions, producing a net gain of 22 resolved instances. To systematically assess the factual reliability of the generated QA, we conduct a human audit on 232 QA pairs drawn from all four transition cells: all pairs from the 44 Fail$\to$Pass and 22 Pass$\to$Fail cases, plus 25 randomly sampled instances each from the Fail$\to$Fail and Pass$\to$Pass cells. Four computer-science master's students, each with four years of development experience, served as reviewers. Each QA pair was independently reviewed by two reviewers, with disagreements resolved through discussion.

Of the 232 audited QA pairs, 230 (99.1\%) are labeled \emph{Supported}: 98 (42.2\%) are fully accurate with all claims grounded in repository evidence, and 132 (56.9\%) contain minor deviations where core claims are correct but local details such as line numbers or function ownership are imprecise, none of which affect the overall conclusion. Only 2 pairs (0.9\%) contain ungrounded claims where the central mechanism assertion is contradicted by the repository code. A detailed breakdown of deviation types and full descriptions of the two ungrounded cases are provided in the supplementary material.

This high factual reliability stems from \approach's QA-driven design: by decomposing a complex issue into targeted, narrowly scoped questions, each \explorertag instance faces a substantially simpler retrieval and reasoning task than answering the full issue at once, reducing the opportunity for hallucination and enabling evidence-grounded answers. The ablation in Section~\ref{sec:rq3_decomp} further confirms this decomposition benefit.

\subsubsection{Influence on Repair Behavior}
\label{sec:stage2_analysis}

To investigate how QA reshapes downstream repair, we compare Mini-SWE-Agent without any pre-repair knowledge (\textbf{w/o QA Injection}) against the same configuration augmented with QA pairs produced by \approach (\textbf{w/ QA Injection}). QA injection reduces the mean number of agent rounds by 7.1\% on the full 500 instances and by 17.1\% on the 44 Fail$\to$Pass instances, confirming that front-loaded repository knowledge bypasses expensive trial-and-error loops~\cite{fan2025sweeffi,bouzenia2025understanding}. The full round-distribution plots and a per-transition API-call shift analysis are provided in the supplementary material.

\begin{figure}[t]
  \centering
  \includegraphics[width=\columnwidth]{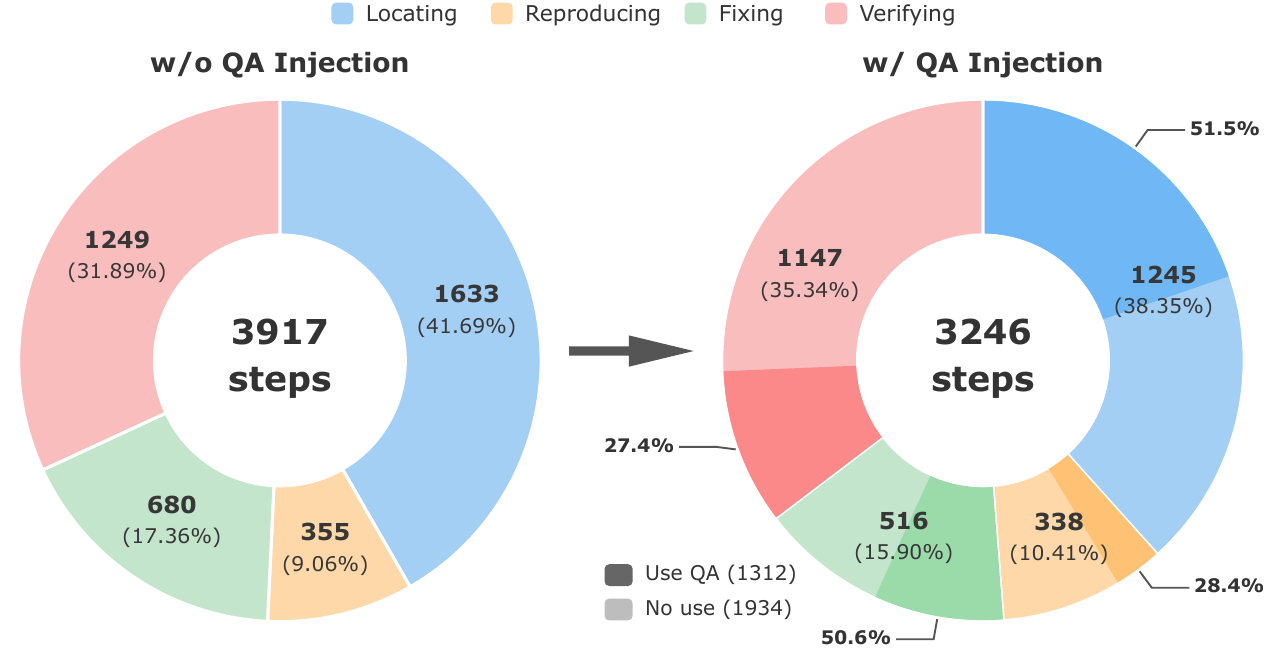}
  \caption{Trajectory stage composition on 44 \emph{Fail$\to$Pass} instances w/o vs.\ w/ QA Injection. Darker shading in the right ring marks QA-related steps.}
  \label{fig:trajectory_stages}
\end{figure}

Since the acceleration is most pronounced on Fail$\to$Pass instances, we focus the remaining trajectory analysis on these 44 cases to study \emph{where} within each trajectory QA knowledge takes effect. We divide each repair trace into four canonical stages~\cite{liu2025empiricalfailures,bouzenia2025understanding,agentless} and annotate whether each step is QA-related, as shown in Figure~\ref{fig:trajectory_stages}. A step is labeled QA-related if the agent explicitly references QA content, or if its action closely follows QA-provided evidence (e.g., opening a file identified in a QA answer or verifying a QA-described behavior). The annotation follows the same protocol described in Section~\ref{sec:rq2}. The total number of repair steps drops from 3{,}917 to 3{,}246, but the reduction is concentrated in the knowledge-intensive phases: Locating contracts by 23.8\% and Fixing by 24.1\%, while Reproducing and Verifying \emph{grow} in proportion (+1.4\,pp and +3.5\,pp). QA knowledge accelerates locating relevant code and composing the fix, freeing the agent to dedicate a larger share of effort to reproduction and verification. This shift parallels the behavior of expert human debuggers, who invest upfront effort in program comprehension and consequently require less trial-and-error exploration~\cite{vessey1985expertise,gugerty1986debugging}.

The step-level annotations confirm this pattern: 40.4\% of all steps are QA-related, with usage concentrated in Locating (51.5\%) and Fixing (50.6\%) compared with Reproducing (28.4\%) and Verifying (27.4\%). The two most accelerated stages are precisely the ones where the agent draws most heavily on QA, providing direct evidence that the speedup is driven by the injected knowledge.

\subsubsection{Pass-to-Fail Regression Analysis}

Although \approach produces a net gain of 22 resolved instances and the preceding analysis demonstrates that QA knowledge substantially reshapes repair behavior, the 22 Pass$\to$Fail regressions warrant closer examination. We manually inspect all 22 regression trajectories and classify each case by whether the injected QA played a misleading role in the repair failure. A case is labeled \emph{misleading} when the dominant downstream effect of the injected QA on the \resolvertag's behavior leads it away from the correct fix.

Under this criterion, only 5 of the 22 regressions are misleading: the QA emphasizes a related but non-gold location or mechanism, and the \resolvertag repeatedly follows this framing during search and editing, eventually deviating from the correct fix. Notably, cross-referencing with the factual audit reveals that among the 10 QA pairs in these 5 cases, only 1 contains an ungrounded claim; the remaining 9 are factually Supported. This indicates that misleading regressions arise primarily from QA providing a plausible but incorrect repair framing, rather than from factual errors in the answers themselves. The remaining 17 regressions are non-misleading: the QA is constructive or neutral, and the failures stem from Resolver-side behavior such as overgeneralizing a correct clue, implementing only part of the required fix, or introducing unnecessary modifications. Detailed behavioral breakdowns for both groups are provided in the supplementary material.

This analysis reveals that the regression bottleneck lies in how the \resolvertag utilizes otherwise reliable QA, not in QA quality itself, suggesting that improving the Resolver's capacity to critically evaluate and selectively apply injected knowledge is a promising direction for reducing regressions.

\begin{findingbox}
\textbf{Finding 2:} The generated QA knowledge is highly reliable, with the vast majority of pairs factually supported by repository evidence. This reliable knowledge accelerates repair by concentrating speedup in the knowledge-intensive Locating and Fixing stages, while the few regressions stem primarily from Resolver-side misuse rather than QA factual errors.
\end{findingbox}

\begin{table}[t]
  \centering
  \small
  \caption{Ablation results on downstream repair performance on SWE-bench Verified under DeepSeek-V3.2.}
  \label{tab:ablation_compare}
  \begin{tabular}{lcc}
  \toprule
  \textbf{Method} & \textbf{Pass@1 (\%)} & \textbf{$\Delta$} \\
  \midrule
  \approach & \textbf{70.8} & -- \\
  \approach-Proposal & 66.0 & \textcolor{red!70!black}{$\downarrow$4.8} \\
  \approach-FreeQ & 67.0 & \textcolor{red!70!black}{$\downarrow$3.8} \\
  \bottomrule
  \end{tabular}
\end{table}

\subsection{RQ3: Ablation on Key Design Choices}
\label{sec:rq3}

To validate the contribution of each key design in \approach, we construct two ablation variants and evaluate them on SWE-bench Verified under DeepSeek-V3.2. Each variant removes one core component while keeping the rest of the pipeline unchanged:

\begin{itemize}[leftmargin=*, itemsep=4pt, topsep=2pt]
  \item \textbf{\approach-Proposal} replaces the \questionertag--\explorertag QA pipeline with a single proposal-generation step. For fairness, the proposal agent shares the same instruction scaffold as \explorertag; only the task differs: it reads the issue and directly produces a solution proposal containing root-cause analysis and suggested fix strategy, instead of answering a targeted repository question. This proposal substitutes for the structured QA pairs injected into \resolvertag.
  
  \item \textbf{\approach-FreeQ} retains the full QA pipeline but removes the category-driven question template from \questionertag. Rather than following the four prescribed question categories, \questionertag freely generates $N$ ($N{=}2$ by default) questions without any categorical constraint.
\end{itemize}

Table~\ref{tab:ablation_compare} reports the results. The following two subsections examine each variant in detail.

\subsubsection{QA Decomposition vs.\ Proposal}
\label{sec:rq3_decomp}
As shown in Table~\ref{tab:ablation_compare}, \approach-Proposal causes the largest performance drop among all ablation variants, reducing Pass@1 by 4.8 percentage points to 66.0\%. Notably, this score even falls below that of Mini-SWE-Agent, i.e., the repair agent without any pre-repair information (66.4\%, Table~\ref{tab:main_results}), meaning the injected proposal not only fails to help but actively \emph{degrades} downstream repair. The root cause is that a single-pass proposal must simultaneously identify root causes, reason about mechanisms, and suggest a fix strategy, producing an analysis that tends to be superficial and entangled when the issue is complex. By contrast, \approach decomposes the issue into narrowly scoped questions, each isolating one specific aspect and allowing the corresponding answer to be reliably grounded through focused exploration. This structured decomposition is a critical enabler of \approach's performance.

\begin{table}[t]
  \centering
  \small
  \caption{Question-quality comparison between \approach-FreeQ and \approach on SWE-bench Verified.}
  \label{tab:rq3_ablation_question_quality}
  \begin{tabular}{l c c c}
  \toprule
  \textbf{Metric} & \textbf{\approach-FreeQ} & \textbf{ACQUIRE} & \textbf{$\Delta$} \\
  \midrule
  Relevance & 8.22 & \textbf{8.49} & \textcolor{green!50!black}{$\uparrow$0.27} \\
  Answerability & \textbf{8.65} & 8.64 & \textcolor{red!70!black}{$\downarrow$0.01} \\
  Diagnostic Utility & 7.58 & \textbf{7.96} & \textcolor{green!50!black}{$\uparrow$0.38} \\
  Reasoning Depth & 6.90 & \textbf{7.28} & \textcolor{green!50!black}{$\uparrow$0.38} \\
  Clarity & 8.42 & \textbf{8.50} & \textcolor{green!50!black}{$\uparrow$0.08} \\
  Coverage & 5.97 & \textbf{6.74} & \textcolor{green!50!black}{$\uparrow$0.78} \\
  \bottomrule
  \end{tabular}
\end{table}

\subsubsection{Category-Guided vs.\ Free Question Generation}
\label{ablation_2}
\noindent
Removing the category-driven template (\approach-FreeQ) also leads to a notable decline, with Pass@1 dropping by 3.8 percentage points to 67.0\%. To understand \emph{why}, we directly analyze question quality using an LLM-as-a-judge protocol~\cite{gu2024survey,shi2025positionbias,tripathi2025pairwise} with DeepSeek-V3.2, evaluating five question-level dimensions and one set-level dimension (\emph{coverage}). Each judgment is scored on a 1--10 scale and averaged over 10 repetitions. Full scoring criteria and evaluation prompts are provided in the supplementary material.

As shown in Table~\ref{tab:rq3_ablation_question_quality}, the category-guided questions achieve the most prominent advantages in \emph{diagnostic utility} (+0.38), \emph{reasoning depth} (+0.38), and \emph{coverage} (+0.78), while \emph{answerability} is virtually identical ($-$0.01). The largest gain in \emph{coverage} confirms that the category-driven template effectively prevents questions from clustering around a single diagnostic angle. A complementary pairwise voting analysis~\cite{shi2025positionbias} further corroborates this: \approach is preferred in 294 of 500 comparisons versus 111 for \approach-FreeQ (95 ties), yielding a non-loss rate of 77.8\%.

\begin{findingbox}
\textbf{Finding 3:} Replacing question decomposition with a single-pass proposal drops Pass@1 by 4.8\,pp, and removing the category-driven template drops it by 3.8\,pp with notably lower question quality, confirming both designs as essential to \approach.

\end{findingbox}

\subsection{RQ4: Sensitivity to the Number of QA Pairs}

We investigate how $N$, the number of QA pairs generated per issue, affects repair performance and cost.

\begin{figure}[t]
  \centering
  \includegraphics[width=\columnwidth]{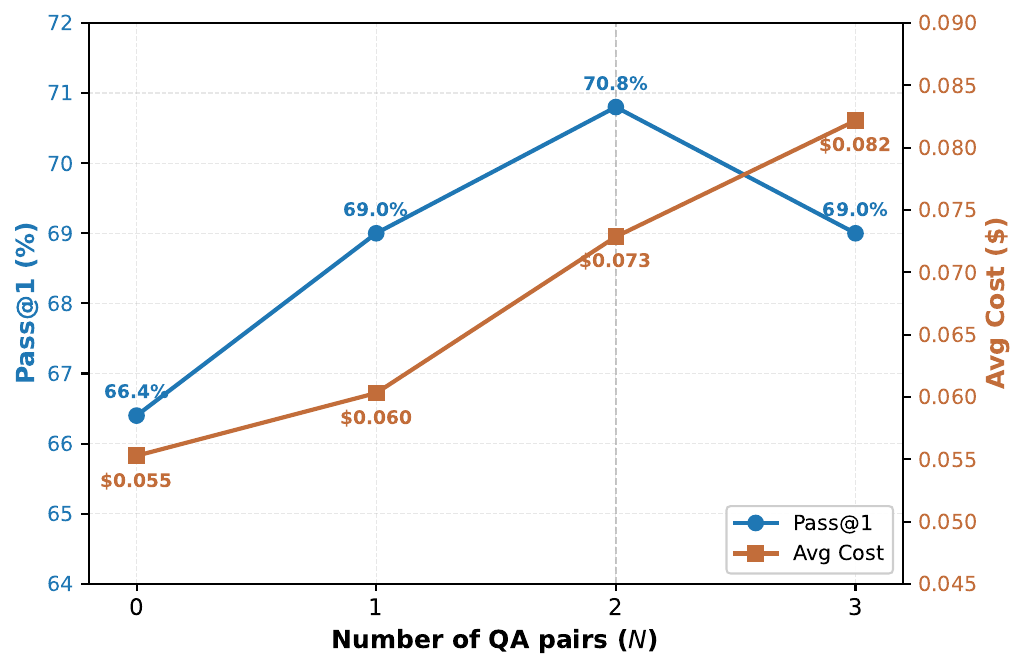}
  \caption{Pass@1 and average cost per instance under varying the number of QA pairs.}
  \label{fig:hyperparameter}
\end{figure}

Figure~\ref{fig:hyperparameter} reports Pass@1 and average cost as functions of $N$. When $N{=}0$, no QA is injected and the pipeline degrades to the Resolver-only baseline (i.e., Mini-SWE-Agent), which achieves 66.4\% Pass@1 at \$0.055 per instance. Introducing even a single QA pair ($N{=}1$) raises Pass@1 by 2.6 percentage points (from 66.4\% to 69.0\%) with only \$0.005 additional cost, demonstrating that the QA-driven paradigm within \approach delivers meaningful improvement under only one pair of QA knowledge. Across all $N{\geq}1$ settings, \approach consistently outperforms every baseline reported in Table~\ref{tab:main_results}, confirming its robustness to the number of QA pairs and highlighting the superiority of semantically rich repository knowledge over alternative forms of repository information such as localization pointers or structural summaries.

Pass@1 peaks at $N{=}2$ (70.8\%) and then drops back to 69.0\% at $N{=}3$, while average cost continues to rise. We attribute this non-monotonic pattern to two factors. First, two questions from complementary categories cover distinct knowledge gaps more effectively than a single question. Second, a third pair tends to overlap with already-covered knowledge, and the longer injected context may dilute the repair agent's attention rather than strengthen it, yielding diminishing returns~\cite{liu2024lost,shi2023large,yang2025distracted}. Based on this cost--performance profile, we adopt $N{=}2$ as the default setting for all other experiments.

\begin{findingbox}
\textbf{Finding 4:} \approach consistently outperforms all baselines across all tested $N{\geq}1$ settings, with performance peaking at $N{=}2$, which offers the best trade-off between effectiveness and cost.
\end{findingbox}

\subsection{Case Study}
\label{sec:case-study}

\begin{figure*}[t]
  \centering
  \includegraphics[width=0.92\textwidth]{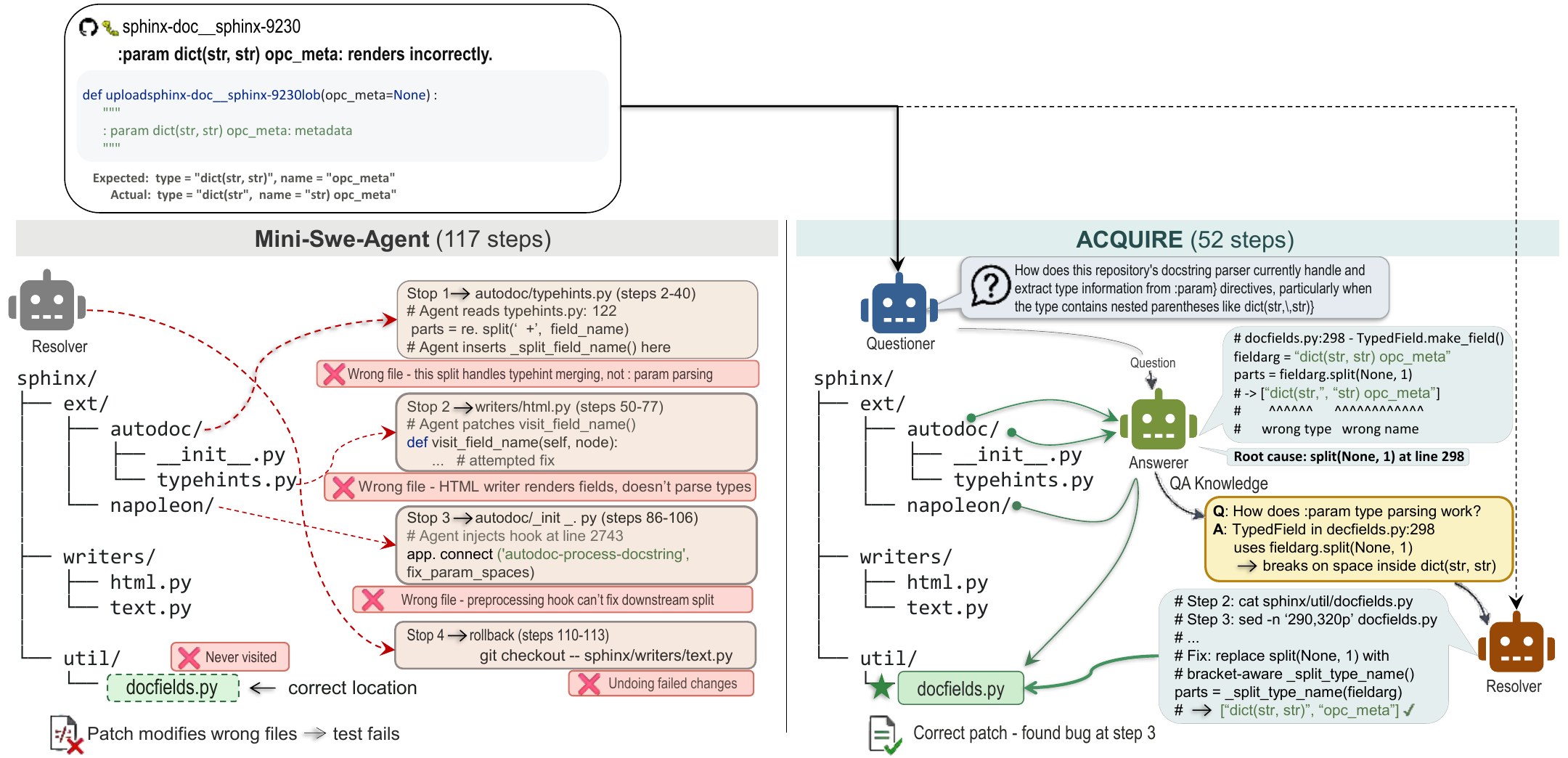}
  \caption{Case Study of \approach with instance sphinx-doc\_\_sphinx-9230.}
  \label{fig:case-study}
\end{figure*}

We illustrate the effectiveness of \approach through a representative example from SWE-bench Verified under DeepSeek-V3.2: \texttt{sphinx-doc\_\_sphinx-9230}
(Figure~\ref{fig:case-study}). The issue reports that Sphinx incorrectly renders \texttt{:param dict(str,\,str) opc\_meta:} because \texttt{fieldarg.split(None,\,1)} in \texttt{sphinx/util/docfields.py} splits on the first whitespace \emph{inside} the parenthesized type, misaligning the type--name boundary.

\vspace{2pt}\noindent\textbf{Mini-SWE-Agent (117 steps).} 
Without pre-repair knowledge, the agent is misled by surface-level keyword overlap. It patches a superficially similar split in \texttt{typehints.py}, then pivots to \texttt{writers/html.py} and \texttt{autodoc/\_\_init\_\_.py} after each attempt fails testing, cycling through increasingly ad-hoc workarounds across four files. The correct file \texttt{docfields.py} is never visited.

\vspace{2pt}\noindent\textbf{ACQUIRE (52 steps).} We describe how each stage of \approach contributes to solving this issue.

\emph{Stage~1: Knowledge Acquisition.}
The \questionertag generates a question targeting the \emph{parsing mechanism} of \texttt{:param} type extraction, rather than the rendering pipeline. This directs exploration toward the causal logic of how types are parsed, enabling discovery of utility-layer code that shares no surface lexical overlap with the issue description. The \explorertag first examines keyword-related modules such as \texttt{autodoc/} and \texttt{napoleon/}, but finds that neither contains the actual \texttt{:param} type-parsing logic. It then broadens its search to \texttt{sphinx/util/} and discovers \texttt{docfields.py}, pinpointing the root cause: the split breaks when spaces appear inside parenthesized types. The answer provides the exact file path, class name, line number, and root-cause explanation.

\emph{Stage~2: Knowledge-Informed Repair.}
With the acquired QA knowledge, the \resolvertag directly opens \texttt{docfields.py}, confirms the problematic \texttt{split} call, and implements a bracket-aware replacement. The resulting patch modifies exactly one file with a minimal, focused change. Notably, the agent designs comprehensive test cases covering diverse compound-typed \texttt{:param} directives to verify the fix, and the final patch directly addresses the root cause identified in the QA answer, echoing both the behavioral shift toward verification and the strong QA--patch alignment observed in Finding~2. The entire trajectory completes in \textbf{52 steps}, a \textbf{55\%} reduction from the baseline's 117~steps.

\vspace{2pt}\noindent\textbf{Analysis.}
Beyond Mini-SWE-Agent, we also inspect the pre-repair exploration outputs of other baselines: none surfaces the root-cause file \texttt{docfields.py}, as the issue description centers on rendering symptoms rather than the underlying parsing logic, misleading keyword-driven exploration away from the actual bug site. This case highlights two advantages of \approach. First, by decomposing the issue into a targeted question, \approach decouples knowledge acquisition from repair and directs the \explorertag to investigate the underlying mechanism rather than chase surface-level keyword matches, yielding a precise and well-grounded answer. Second, the semantically rich QA knowledge allows the repair agent to inherit accurate fault localization and root-cause understanding, effectively offloading a significant portion of the exploratory burden and substantially improving both efficiency and repair capability.

\section{Discussion}
\subsection{Strengths}

Our study demonstrates that explicitly acquiring repository knowledge before repair is practical and effective for repository-level issue resolution. By decoupling question generation, repository-grounded answering, and patch synthesis, \approach turns implicit knowledge gaps into explicit QA knowledge before editing begins. This design directly addresses the recurring knowledge-deficit failure patterns identified in Section~\ref{sec:motivation}, and operationalizes a principle long observed in empirical debugging research: expert developers consistently invest more effort in program comprehension before attempting repairs, yielding faster and more accurate outcomes than jumping directly into code modification~\cite{vessey1985expertise,gugerty1986debugging}.

Empirically, \approach shows consistent gains across different backbone models at modest additional cost. The improvements are not only reflected in Pass@1, but also in behavioral efficiency: the repair agent spends less effort on blind locating and trial-and-error fixing, and more on reproduction and verification. This indicates that pre-repair knowledge acquisition improves both effectiveness and the quality of repair trajectories.

Another strength is interpretability. The category-guided question template provides a clear interface for what knowledge is being requested, while the QA pairs make it easier to inspect, diagnose, and analyze why a repair succeeds or fails. Compared with the localization lists or repair descriptions produced by existing pre-repair methods, this structured QA pipeline offers better controllability and clearer debugging signals for future system improvement.

\subsection{Limitations and Future Work}
Despite these strengths, the current framework still has room for further improvement.

The four-category question template, while practical and effective, can be further refined for broader coverage. Future work can explore hierarchical categories, issue-type-adaptive templates, or automatically learned category schemas from large-scale repair trajectories.

Additionally, knowledge is currently injected as pre-generated QA context before repair, which has proven simple and stable in practice. A promising extension is to couple knowledge acquisition with the repair trajectory itself, allowing the agent to dynamically request and refresh knowledge as new hypotheses emerge during debugging. However, dynamic acquisition would inevitably introduce additional cost; exploring how to balance knowledge freshness and computational cost remains an open challenge.

\section{Threats to Validity}

We discuss potential validity threats from three perspectives.

\textbf{Construct validity.}
A key threat is whether our metrics fully capture repair quality. We use Pass@1 on SWE-bench Verified as the primary effectiveness metric, and complement it with time, cost, and trajectory-step analyses that provide an efficiency perspective on the repair process.

\textbf{Internal validity.}
Observed gains could be influenced by confounding factors such as prompt design, tool behavior, or hyperparameter choices rather than the proposed QA-driven mechanism itself. To reduce this risk, we keep the repair backbone and evaluation protocol consistent across methods, and conduct targeted ablations that remove core components (question decomposition and category-guided question generation). The consistent performance drops in these ablations support the causal contribution of our key design choices. Additionally, the question-quality evaluation in RQ3 relies on LLM-as-a-judge; to mitigate potential scoring bias, we adopt both scoring and pairwise voting perspectives, repeat each judgment 10 times, and apply position-bias mitigation following established practices~\cite{shi2025positionbias}.

\textbf{External validity.}
Our current prompt design and evaluation are tailored to Python repositories. While the QA-driven paradigm itself is language-agnostic, adapting the prompt design to other programming languages and validating its effectiveness across them remains future work.

\section{Related Work}

\subsection{Software Issue Resolution}

Repository-level issue resolution addresses real-world software bugs by reasoning over cross-file dependencies and generating patches within full repositories. Unlike function-level repair, it demands broader context understanding and robust handling of ambiguous descriptions. Recent progress is driven by the SWE-bench benchmarks~\cite{jimenez2023swe,openai2024swebenchverified,ma2026llmagentscoderepositories} and supporting infrastructure such as Repo2Run for scalable environment construction~\cite{hu2025repo2run}, while agentic frameworks like SWE-agent enable models to interact with realistic software environments~\cite{yang2024sweagenta}. Alongside these advances, a growing body of work scrutinizes benchmark reliability, examining issues such as test overfitting, label quality, evaluation inflation, and benchmark extensibility~\cite{wang2025solvedissues,ahmed2025testoverfitting,yu2026sweabs,oliva2025spice,chen2025oldtests,yu2025utboost,wang2025swebenchplusplus,garg2025savingswebench}.

Existing approaches follow three main directions. First, structured workflow decomposition improves controllability and reasoning. For instance, Agentless splits repair into localization, repair, and validation stages~\cite{agentless}. AutoCodeRover and LingmaAgent enhance repository search and multi-step reasoning~\cite{autocoderover,lingmaagent}, while SWE-Debate introduces structured debate into the repair pipeline~\cite{li2025swe}. 

Second, effective repository exploration and context management are critical. LocAgent uses graph-guided multi-hop reasoning for localization~\cite{locagent}, and RepoGraph leverages structural representations for better traversal~\cite{repograph}. Additionally, EET and SWE-Pruner emphasize efficiency control and context pruning to streamline search and reading processes~\cite{guo2026eet,wang2026swe}, and recent work further explores compressing code context to reduce input length while preserving repair-relevant information~\cite{jia2026compressing}. 

Third, leveraging reusable knowledge and memory enhances agentic software engineering~\cite{lin2024llms}. Tools like SWE-Exp, EXPEREPAIR, and Agent KB utilize historical experience, dual-memory designs, and shared knowledge bases to inform repair decisions~\cite{chen2025swe,mu2025experepair,agentkb}. Similarly, SAGE and SE-Agent reuse execution trajectories for self-evolution and planning~\cite{hayashi2025self,lin2025se}.

Empirical studies further illuminate systematic failure patterns in automated issue resolution~\cite{liu2025empiricalfailures,bouzenia2025understanding,ma2026samesignal}. These analyses identify knowledge deficits, particularly insufficient understanding of repository internals and behavioral contracts, as a primary source of incorrect or incomplete patches, motivating explicit pre-repair knowledge acquisition. Despite these advances, prior methods assume that existing mechanisms can surface all relevant evidence, neglecting explicit knowledge-gap identification before patch synthesis~\cite{yang2024sweagenta,agentless,autocoderover,lingmaagent,li2025swe,locagent,repograph,guo2026eet,wang2026swe,chen2025swe,mu2025experepair,hayashi2025self,lin2025se,agentkb}. \approach overcomes this limitation by decoupling question generation, repository-grounded answering, and patch synthesis, enabling proactive acquisition of missing contextual evidence prior to repair.

\subsection{Repository QA and Knowledge Acquisition}

Repository-level QA has evolved from semantic code search benchmarks such as CodeSearchNet and CoSQA+, which align natural-language intents with code snippets~\cite{codesearchnet,cosqaplus}. Modern efforts scale this to full repositories, requiring multi-file traversal and evidence aggregation across multiple interdependent files.

Recent benchmarks formalize repository-scale comprehension. RepoQA evaluates long-context code understanding~\cite{repoqa}, while CodeRepoQA, CoReQA, and SWE-QA introduce large-scale QA datasets derived from software engineering tasks, GitHub issues, and complex multi-hop dependencies~\cite{coderepoqa,coreqa,sweqa}. These benchmarks reveal that repository-level QA demands not only retrieval precision but also the ability to synthesize answers from evidence scattered across multiple abstraction layers.

Repository QA increasingly intersects with agentic reasoning and structured retrieval. Tools like RepoGraph enhance navigation via structural representations~\cite{repograph}, while SWE-agent and RepoCoder utilize iterative retrieval and generation for interactions and code completion~\cite{yang2024sweagenta,repocoder}. Furthermore, RepoQA-Agent shifts repository QA from static retrieval to interactive, search-driven answering guided by reinforcement learning~\cite{repoqa_agent}.

Despite these advancements, existing QA frameworks typically assume a human-to-repository paradigm where the question is pre-provided, focusing solely on retrieval and answer generation~\cite{repoqa,coderepoqa,coreqa,sweqa,repograph,repoqa_agent}. \approach shifts this to an agent-to-repository model. Rather than treating QA as an external benchmark, \approach integrates it as an internal capability, enabling the agent to autonomously identify knowledge gaps, formulate questions, and acquire contextual evidence prior to patch generation.

\section{Conclusion}

This paper presented \approach, a QA-driven framework for repository-level software issue resolution. Instead of directly searching and patching from issue keywords, \approach explicitly identifies missing knowledge, acquires repository-grounded answers, and then performs repair with this structured evidence.

Across experiments on SWE-bench Verified, \approach consistently improves issue resolution over strong baselines at modest additional cost and remains robust under different model backbones. Behavioral analysis further reveals that QA injection accelerates knowledge-intensive repair stages and shifts agent effort toward verification, with the acquired knowledge ultimately encoded into the majority of successful patches. Ablation studies confirm that the gains stem from two key design choices: decomposing issues into targeted, answerable questions and using category-guided question generation to obtain more diagnostic and complementary knowledge.

Overall, our results suggest that explicit pre-repair knowledge acquisition is a critical ingredient for reliable agentic software engineering, and \approach provides a practical path toward more accurate, efficient, and interpretable automated issue resolution.

\section*{Data Availability}

All code and data used in this study are publicly available at: \url{https://github.com/LionLin2003/ACQUIRE}.

\bibliographystyle{IEEEtran}
\bibliography{ref}

\clearpage
\setcounter{section}{0}
\setcounter{subsection}{0}
\setcounter{subsubsection}{0}
\setcounter{figure}{0}
\setcounter{table}{0}

\twocolumn[
\begin{@twocolumnfalse}
\begin{center}
{\LARGE\bfseries Supplementary Material\par}
\vspace{1.0em}
\end{center}
\end{@twocolumnfalse}
]

\renewcommand{\dbltopfraction}{0.95}
\renewcommand{\dblfloatpagefraction}{0.95}
\renewcommand{\textfraction}{0.05}

\section{Prompt Templates}
\label{appendix:prompts}

\subsection{Questioner Prompt}
\label{appendix:prompt-questioner}

\begin{tcblisting}{promptbox={System Prompt}}
You are an expert software engineer tasked with generating questions to help resolve GitHub issues.

# Task Overview
Generate exactly {N} questions that are most relevant to resolving the given issue. These questions should focus on understanding the repository's current state and implementation.

# Question Categories
Select from the following 4 categories based on the issue content. Categories can be repeated if both top questions fall into the same category.

## 1. Mechanism & Behavior
Questions about specific functional logic flows, state management, data processing details, and internal workings of specific components. Questions that ask "**how does this functionality actually work**".

## 2. Design & Usage
Questions about API definitions, class inheritance hierarchies, error specifications, design pattern applications, and type safety design. Questions that ask "**how interfaces are designed and what contracts they follow**".

## 3. Locating & Structure
Questions about the codebase's macro layout, module division, dependency relationships, and file location knowledge. Questions that ask "**where things are and what the overall structure looks like**".

## 4. Ecosystem & Standards
Questions about external technical ecosystems and general rules that the project depends on, including third-party libraries, programming language features, official standard libraries, industry protocols, file format standards, and mathematical/scientific concepts. Questions that ask "**what external knowledge is needed**".

Note: Categories 1-3 are repository knowledge that can be obtained from within the project. Category 4 is external knowledge independent of the repository.

# Guidelines

## DO:
- Focus on understanding the repository's current implementation
- Ask questions whose answers can be obtained by analyzing the code repository
- Ask about the current state of the code as it exists now
- Prioritize questions by importance (Q1 is most important, Q2 is second)
- Use generic terms like "this repository", "the codebase", "this project" instead of specific project names

## DO NOT:
- Ask about issue-specific details that are only in the issue description
- Ask about external knowledge or resources outside the repository (unless for Category 4)
- Ask about code changes, git operations, commits, branches, or versions
- **NEVER reference specific commit hashes, commit numbers, or version changes**
- **NEVER ask about "changes introduced in commit X" or "what was modified in PR Y"**
- **NEVER use specific project names - use "this repository", "the codebase", "this project" instead**
- Generate duplicate or highly similar questions

# Output Format
Return ONLY a valid JSON object with the following structure:

```json
{
  "Q1": {
    "category": "Mechanism & Behavior",
    "question": "How does the authentication system validate user credentials in this repository?"
  },
  "Q2": {
    "category": "Locating & Structure",
    "question": "Where is the login functionality implemented in the codebase?"
  }
}
```

Important:
- Generate exactly {N} questions
- Questions should be sorted by importance (Q1 being most important)
- Each question must have both "category" and "question" fields
- Use the exact category names provided above
- Categories can repeat if both questions belong to the same category
- Do not include any text outside the JSON object
\end{tcblisting}

\begin{tcblisting}{promptbox={User Prompt}}
Here is the GitHub issue problem statement:

{problem_statement}

Please generate exactly {N} questions based on the above problem statement according to the specified requirements and output format.

\end{tcblisting}

\subsection{Answerer Instance Template}
\label{appendix:prompt-answerer}

\begin{tcblisting}{promptbox={Instance Template}}
<question>
{{task}}
</question>

<instructions>
# Task Instructions

## Overview
You're a software engineer analyzing a codebase by interacting with a computer shell.
Your task is to understand the repository and answer the question provided above.

IMPORTANT: This is a READ-ONLY task. You should NOT modify any files in the repository.
Your goal is to explore, read, and analyze the code to answer the question accurately.

For each response:
1. Include a THOUGHT section explaining your reasoning and what you're trying to accomplish
2. Provide exactly ONE bash command to execute

## Important Rules
- READ ONLY: You can read files, search code, list directories, run analysis commands
- DO NOT MODIFY: Do not create, edit, or delete any files in /testbed
- DO NOT RUN: Avoid running tests or executing application code unless necessary for understanding

## Recommended Workflow
1. Understand the question thoroughly
2. Explore the repository structure to locate relevant code
3. Read and analyze relevant files
4. Search for patterns, function definitions, or specific implementations
5. Synthesize information to form your answer
6. Submit your final answer

## Command Execution Rules
You are operating in an environment where:
1. You write a single command
2. The system executes that command in a subshell
3. You see the result
4. You write your next command

Each response should include:
1. A **THOUGHT** section where you explain your reasoning and plan
2. A single bash code block with your command

Format your responses like this:

<format_example>
THOUGHT: Here I explain my reasoning process, analysis of the current situation,
and what I'm trying to accomplish with the command below.

```bash
your_command_here
```
</format_example>

Commands must be specified in a single bash code block:

```bash
your_command_here
```

**CRITICAL REQUIREMENTS:**
- Your response SHOULD include a THOUGHT section explaining your reasoning
- Your response MUST include EXACTLY ONE bash code block
- This bash block MUST contain EXACTLY ONE command (or a set of commands connected with && or ||)
- If you include zero or multiple bash blocks, or no command at all, YOUR RESPONSE WILL FAIL
- Do NOT try to run multiple independent commands in separate blocks in one response
- Directory or environment variable changes are not persistent. Every action is executed in a new subshell.
- However, you can prefix any action with `MY_ENV_VAR=MY_VALUE cd /path/to/working/dir && ...` or write/load environment variables from files

Example of a CORRECT response:
<example_response>
THOUGHT: I need to understand the structure of the repository first. Let me check what files are in the current directory to get a better understanding of the codebase.

```bash
ls -la
```
</example_response>

Example of an INCORRECT response:
<example_response>
THOUGHT: I need to examine the codebase and then look at a specific file. I'll run multiple commands to do this.

```bash
ls -la
```

Now I'll read the file:

```bash
cat file.txt
```
</example_response>

If you need to run multiple commands, either:
1. Combine them in one block using && or ||
```bash
command1 && command2 || echo "Error occurred"
```

2. Wait for the first command to complete, see its output, then issue the next command in your following response.

## Environment Details
- You have a full Linux shell environment
- Always use non-interactive flags (-y, -f) for commands
- Avoid interactive tools like vi, nano, or any that require user input
- All standard Unix tools are available (grep, find, cat, head, tail, etc.)

## Useful Command Examples for Code Analysis

### List repository structure:
```bash
find /testbed -type f -name "*.py" | head -20
```

### Search for specific code patterns:
```bash
grep -r "function_name" /testbed --include="*.py"
```

### View file content:
```bash
cat /testbed/path/to/file.py
```

### View specific lines with numbers:
```bash
nl -ba filename.py | sed -n '10,20p'
```

### Search with context:
```bash
grep -A 5 -B 5 "pattern" filename.py
```

### Find file by name:
```bash
find /testbed -name "models.py"
```

### Count lines or analyze code:
```bash
wc -l /testbed/**/*.py
```

## Answer Submission
When you have gathered enough information and are ready to provide your final answer,
issue exactly the following command:

```bash
cat <<'ANSWER_EOF'
SUBMIT_ANSWER
Your detailed answer here. Be specific, thorough, and reference the code/files you analyzed.
Include relevant file paths, function names, and explanations.
ANSWER_EOF
```

The format is CRITICAL:
- First line after opening must be: SUBMIT_ANSWER
- Following lines contain your answer
- Use the heredoc syntax exactly as shown above

This command will submit your answer.
You cannot continue working on this task after submitting.
</instructions>
\end{tcblisting}

\subsection{Question-Quality Evaluation Prompts}
\label{appendix:prompt-eval}

\subsubsection{Scoring -- Individual Question}

\begin{tcblisting}{promptbox={System Prompt}}
You are a rigorous evaluator assessing the quality of a single repository-focused diagnostic question generated from a GitHub issue.

Evaluate the question only using the provided problem statement and the question text itself.

A strong question should be tightly grounded in the repository, highly relevant to resolving the issue, diagnostically useful for narrowing the debugging search space, and clearly phrased.
\end{tcblisting}

\begin{tcblisting}{promptbox={User Prompt}}
## Input

Problem Statement:
{problem_statement}

Question to Evaluate:
{question}

## Task

Evaluate the single question across the five dimensions below.
Use integer scores from 1 to 10.

## Scoring Dimensions

1. Relevance
- 1-2: Irrelevant to the issue's core failure or likely fix path.
- 3-4: Weakly related but mostly about side details or symptoms.
- 5-6: Moderately relevant but still broad or indirect.
- 7-8: Strongly relevant to understanding or resolving the issue.
- 9-10: Precisely targets the likely failure mechanism, root cause, or key repair decision.

2. Repository Answerability
- 1-2: Requires mainly outside knowledge or general internet or documentation lookup.
- 3-4: Only loosely tied to the repository; answerable mostly with generic software reasoning.
- 5-6: Partly grounded in repository artifacts but still broad.
- 7-8: Clearly answerable through repository code, tests, config, or docs.
- 9-10: Strongly grounded in specific internal implementation, control flow, state propagation, validation logic, or repository-specific architecture.

3. Diagnostic Utility
- 1-2: Adds almost no debugging value.
- 3-4: Some value, but leaves the search space very large.
- 5-6: Moderately useful for narrowing the investigation.
- 7-8: Strongly helps isolate a subsystem, state transition, boundary condition, or execution path.
- 9-10: Highly discriminative; answering it would sharply narrow the fault to a specific mechanism, guard condition, state transition, or repair decision.

4. Reasoning Depth
- 1-2: Trivial lookup or superficial restatement.
- 3-4: Simple one-file or one-symbol lookup.
- 5-6: Requires some local reasoning over behavior or flow.
- 7-8: Requires tracing meaningful interactions across functions, files, or states.
- 9-10: Requires substantial multi-hop reasoning across decoupled components, state changes, or abstraction boundaries.

5. Clarity
- 1-2: Ambiguous, confusing, or combines multiple unrelated asks.
- 3-4: Understandable but poorly scoped or imprecise.
- 5-6: Acceptable but somewhat verbose or fuzzy.
- 7-8: Clear, well-scoped, and technically precise.
- 9-10: Very concise, unambiguous, and sharply targeted.

## Hard Flags

Set each to true or false.

- requires_external_knowledge: answering the question fundamentally depends on external domain knowledge rather than repository artifacts.
- too_generic: the question is generic enough to fit almost any software issue.
- rephrases_issue_without_added_value: the question mostly restates the issue without adding diagnostic direction.

## Guidance

Repository Answerability is about where the answer comes from.
Diagnostic Utility is about how much the answer would reduce debugging uncertainty.
Do not conflate them.

A question that could be answered reasonably well without opening the repository should not receive a high Repository Answerability score.
A question that merely suggests where to start looking, without meaningfully reducing uncertainty, should not receive a high Diagnostic Utility score.

Do not reward complexity for its own sake.
A focused question may deserve a high score even if it does not require broad multi-file reasoning.

## Calibration Examples

Example Good:
- Problem: "Parser crashes when optional config is missing."
- Question: "Where is the default config object created, and which call path assumes it is non-null before validation?"

Example Bad:
- Problem: "Build fails on Python 3.12."
- Question: "What are the exact reproduction steps and environment details?"

Example Weak Repository Question:
- Problem: "A validation error appears in one edge case."
- Question: "Why does this issue happen in the application?"
- Why weak: broad, generic, and not clearly answerable from repository internals.

## Output Format

Return only valid JSON.

{
  "scores": {
    "relevance": 1,
    "repository_answerability": 1,
    "diagnostic_utility": 1,
    "reasoning_depth": 1,
    "clarity": 1
  },
  "hard_flags": {
    "requires_external_knowledge": false,
    "too_generic": false,
    "rephrases_issue_without_added_value": false
  },
  "summary": "One concise sentence explaining the judgment.",
  "dimension_rationale": {
    "relevance": "Short explanation.",
    "repository_answerability": "Short explanation.",
    "diagnostic_utility": "Short explanation.",
    "reasoning_depth": "Short explanation.",
    "clarity": "Short explanation."
  }
}
\end{tcblisting}

\newpage
\subsubsection{Scoring -- Question Set (Coverage)}

\begin{tcblisting}{promptbox={System Prompt}}
You are an expert evaluator assessing the quality of a set of repository-focused diagnostic questions generated from a GitHub issue.

Evaluate the set only using the provided problem statement and the question texts themselves.

A strong question set should efficiently cover the key diagnostic angles needed to understand, localize, and resolve the issue.
High-quality coverage means the questions are collectively relevant to the core defect, complementary rather than repetitive, and useful for narrowing the debugging search space.
\end{tcblisting}

\begin{tcblisting}{promptbox={User Prompt}}
## Input

Problem Statement:
{problem_statement}

Questions:
{question_set}

## Task

Evaluate the question set as a whole using a single dimension: Coverage.
Use an integer score from 1 to 10, where 10 is always the best score.

## Scoring Dimension

1. Coverage
Coverage measures how well this set uses its limited question slots to cover the key diagnostic angles needed to debug this specific issue.

When scoring coverage, consider all of the following together:
- Core relevance: Are the questions collectively centered on the main failure mechanism rather than peripheral details?
- Complementarity: Do the questions probe different useful diagnostic angles rather than circling the same point?
- Non-redundancy: Does each question add distinct information value, instead of rephrasing another question?
- Debugging usefulness: Would answering this set significantly reduce uncertainty about where and why the bug occurs?

Behavioral anchors:
- 1-2: Very poor coverage. The set is generic, repetitive, off-target, or misses the key diagnostic needs of the issue.
- 3-4: Weak coverage. Some questions are related, but the set leaves major diagnostic gaps and wastes question slots on overlap or shallow prompts.
- 5-6: Moderate coverage. The set addresses part of the debugging space, but misses important complementary angles or contains noticeable redundancy.
- 7-8: Strong coverage. The set covers several important and complementary diagnostic angles with mostly useful and non-redundant questions.
- 9-10: Excellent coverage. The set efficiently covers the most important diagnostic angles needed for this issue, with highly complementary, targeted, and information-dense questions.

## Guidance

Do not reward breadth or question count for their own sake.
A smaller set can score higher if it covers the key diagnostic needs more efficiently.
Reworded versions of the same underlying question should not be treated as additional coverage.
Coverage should reflect usefulness for debugging this specific issue, not general interestingness.

## Calibration Examples

Example Better Set:
- Q1 identifies the likely failing file, subsystem, or execution path.
- Q2 asks how the relevant control flow or state transition currently works.
- Why better: covers complementary diagnostic angles with strong debugging value and little redundancy.

Example Worse Set:
- Q1 asks where the bug is implemented.
- Q2 asks which file contains the bug logic.
- Why worse: repeated intent, weak complementarity, and poor effective coverage.

## Output Format

Return only valid JSON.

{
  "scores": {
    "coverage": 1
  },
  "summary": "One concise sentence explaining the judgment.",
  "dimension_rationale": {
    "coverage": "Short explanation referencing relevance, complementarity, redundancy, and debugging usefulness."
  }
}
\end{tcblisting}

\subsubsection{Voting}

\begin{tcblisting}{promptbox={System Prompt}}
You are a rigorous evaluator comparing two candidate sets of repository-focused diagnostic questions generated for the same GitHub issue.

Evaluate the two sets only using the provided problem statement and the question texts.

Your goal is to decide which set is overall more useful for helping a developer understand, localize, and resolve the issue.

Bias controls:
- Ignore presentation order. Evaluate only by merit.
- Do not reward verbosity, larger set size, or more complex language by default.
- Prefer concise, high-signal, repository-grounded questions over bloated or generic ones.
- If the two sets are functionally equivalent for debugging, return TIE instead of forcing a winner.
\end{tcblisting}

\begin{tcblisting}{promptbox={User Prompt}}
## Input

Problem Statement:
{problem_statement}

Set A Questions:
{questions_a}

Set B Questions:
{questions_b}

## Task

Compare Set A and Set B holistically and choose the better set.

## Decision Criteria

Compare the two sets across these dimensions:

1. Relevance
- Which set better targets the core failure mechanism or likely repair path?

2. Repository Answerability
- Which set is more clearly answerable from repository code, tests, config, or docs rather than generic software reasoning or outside knowledge?

3. Diagnostic Utility
- Which set would do more to reduce debugging uncertainty and narrow the search space?

4. Reasoning Depth
- Which set more often asks for meaningful behavioral or structural reasoning rather than superficial lookup?

5. Coverage and Efficiency
- Which set better covers complementary diagnostic angles while minimizing semantic overlap and wasted question slots?

## Guidance

Repository Answerability is about where the answer comes from.
Diagnostic Utility is about how much the answer would help debugging.
Do not conflate them.

Do not reward complexity for its own sake.
A simpler but sharper set should beat a more verbose but less useful set.

Do not reward a set just because it has more questions.
Better coverage means stronger complementary value, not larger quantity.

Return TIE when the practical debugging benefit is effectively the same.

## Calibration Examples

Example Better Set:
- One question asks where the relevant subsystem is implemented.
- One asks how the failing control flow or state transition currently behaves.
- One asks about a boundary condition, validation path, or missing test.
- Why better: strong complementarity, high debugging value, low redundancy.

Example Worse Set:
- Multiple questions all ask where the bug is located, which file is responsible, or where the logic lives.
- Why worse: repetitive, weakly complementary, wastes question slots.

## Output Format

Return only valid JSON.

{
  "winner": "A",
  "confidence": 1,
  "summary": "One or two concise sentences explaining the decisive difference."
}

Notes:
- winner must be one of: A, B, TIE
- confidence must be an integer from 1 to 10
\end{tcblisting}

\clearpage

\section{Supplementary Experimental Analysis}
\label{appendix:exp_analysis}

\subsection{Stage-wise Cost and Time Analysis}
\label{appendix:cost_time}

Table~\ref{tab:main_results} in the main paper reports the aggregate average time per instance. Table~\ref{tab:cost_time} provides a finer-grained breakdown, separating the pre-repair exploration stage from the downstream repair stage for both wall-clock time and monetary cost.

\begin{table}[H]
  \caption{Stage-wise time and cost per instance.}
  \centering
  \scriptsize
  \setlength{\tabcolsep}{2.5pt}
  \renewcommand{\arraystretch}{0.92}
  \resizebox{\columnwidth}{!}{%
  \begin{tabular}{@{}ll rrr rrr@{}}
    \toprule
    & & \multicolumn{3}{c}{\textbf{Time (s)}}
      & \multicolumn{3}{c}{\textbf{Cost (\$)}} \\
    \cmidrule(lr){3-5} \cmidrule(lr){6-8}
    \textbf{Method} & \textbf{Model}
      & \textbf{Pre} & \textbf{Repair} & \textbf{Total}
      & \textbf{Pre} & \textbf{Repair} & \textbf{Total} \\
    \midrule
    \multirow{2}{*}{Mini-SWE-Agent} & GPT-5-mini
      & 0 & 187 & 187
      & 0.000 & 0.024 & 0.024 \\
    & DeepSeek-V3.2
      & 0 & 815 & 815
      & 0.000 & 0.055 & 0.055 \\
    \midrule
    \multirow{2}{*}{LocAgent} & GPT-5-mini
      & 106 & 331 & 437
      & 0.023 & 0.025 & 0.048 \\
    & DeepSeek-V3.2
      & 167 & 879 & 1046
      & 0.021 & 0.038 & 0.060 \\
    \midrule
    \multirow{2}{*}{CoSIL} & GPT-5-mini
      & 42 & 182 & 224
      & 0.007 & 0.028 & 0.035 \\
    & DeepSeek-V3.2
      & 31 & 719 & 750
      & 0.004 & 0.055 & 0.059 \\
    \midrule
    \multirow{2}{*}{LingmaAgent} & GPT-5-mini
      & 651 & 171 & 823
      & 0.286 & 0.022 & 0.309 \\
    & DeepSeek-V3.2
      & 640 & 781 & 1421
      & 0.106 & 0.049 & 0.155 \\
    \midrule
    \multirow{2}{*}{SWE-Debate} & GPT-5-mini
      & 1392 & 160 & 1552
      & 0.714 & 0.024 & 0.738 \\
    & DeepSeek-V3.2
      & 1675 & 843 & 2517
      & 0.329 & 0.053 & 0.382 \\
    \midrule
    \multirow{2}{*}{\textbf{\approach}} & GPT-5-mini
      & 125 & 177 & 302
      & 0.027 & 0.026 & 0.054 \\
    & DeepSeek-V3.2
      & 259 & 783 & 1042
      & 0.022 & 0.051 & 0.073 \\
    \bottomrule
  \end{tabular}
  }
  \label{tab:cost_time}
\vspace{-0.6em}
\begin{flushleft}
\scriptsize \emph{Note.} Results are measured on SWE-bench Verified
(500 instances). ``Pre'' denotes the pre-repair stage; ``Repair''
denotes the downstream repair agent.
\end{flushleft}
\vspace{-1.0em}
\end{table}

The stage-wise breakdown reveals several findings. First, compared with LingmaAgent and SWE-Debate, \approach's pre-repair stage is 2.5--11$\times$ faster and 5--26$\times$ cheaper, while achieving the best Pass@1. Second, compared with LocAgent and CoSIL, \approach's pre-repair overhead is moderate, and total cost stays comparable, while delivering a clearly higher Pass@1. Third, \approach adds only modest end-to-end overhead over the bare Mini-SWE-Agent ({$\sim$}115\,s\,/\,{$\sim$}227\,s and \$0.030\,/\,\$0.018 per instance) for the +3.8\,/\,+4.4 Pass@1 gain. In addition, since the $N$ \explorertag instances are independent, the \explorertag stage can be parallelized, which would further compress wall-clock time without changing total computation.

\subsection{Agent Round and API-call Analysis}
\label{appendix:round_analysis}

\subsubsection{Agent Round Distribution}

\begin{figure}[H]
  \centering
  \includegraphics[width=\columnwidth]{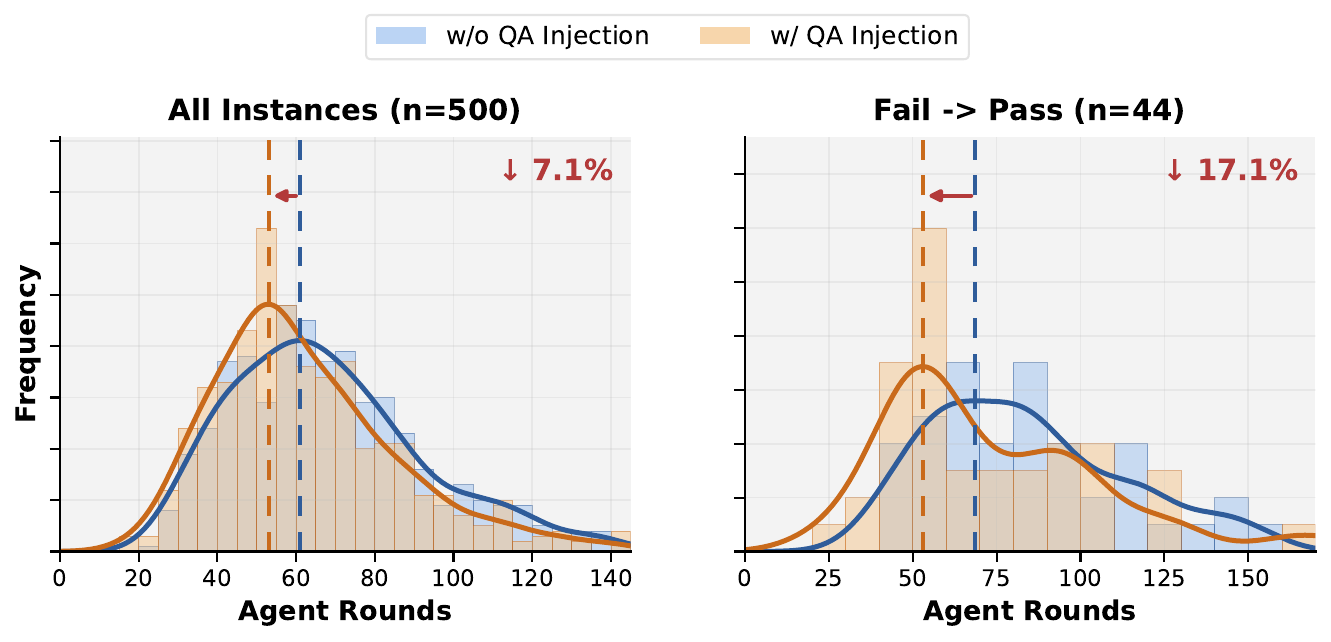}
  \caption{Agent round distribution w/o vs.\ w/ QA injection under DeepSeek-V3.2.}
  \label{fig:api_distribution}
\end{figure}

Figure~\ref{fig:api_distribution} plots the distribution of agent rounds with and without QA injection under DeepSeek-V3.2. On the full set of 500 instances (left), QA injection shifts the distribution leftward, reducing the mean number of rounds by 7.1\%. The shift becomes far more pronounced on the 44 Fail$\to$Pass instances (right), where the mean reduction reaches 17.1\%. This confirms that QA injection directly mitigates the disproportionate cost of knowledge-deficient repair attempts~\cite{fan2025sweeffi,bouzenia2025understanding}, by front-loading the missing repository understanding and bypassing expensive trial-and-error loops.

\subsubsection{API-call Shift by Transition Type}

To further examine how QA injection affects exploration effort across different outcome groups, we compute the signed difference in the number of API calls between \approach and Mini-SWE-Agent ($\Delta = \text{API}_{\approach} - \text{API}_{\text{Mini-SWE-Agent}}$) for each instance and report the results grouped by transition type in Table~\ref{tab:api_call_shift}.

\begin{table}[H]
  \caption{API-call shift between Mini-SWE-Agent and \approach on SWE-bench Verified under DeepSeek-V3.2.}
  \centering
  \small
  \setlength{\tabcolsep}{4pt}
  \begin{tabular}{@{}l r r r r r@{}}
    \toprule
    \textbf{Group} & \textbf{\#Cases} & \textbf{Decrease}
      & \textbf{Same} & \textbf{Increase}
      & \textbf{Mean $\Delta$} \\
    \midrule
    Fail$\to$Pass              & 44 & 32 & 0 & 12 & $-$15.18 \\
    Pass$\to$Fail overall      & 22 & 10 & 0 & 12 & $+$0.55  \\
    \quad Misleading            &  5 &  1 & 0 &  4 & $+$11.40 \\
    \quad Non-misleading        & 17 &  9 & 0 &  8 & $-$3.64  \\
    \bottomrule
  \end{tabular}
  \label{tab:api_call_shift}
\end{table}

Fail$\to$Pass cases show a large reduction in API calls (mean $\Delta = -15.18$), consistent with QA reducing trial-and-error exploration. The aggregate Pass$\to$Fail increase is small ($+0.55$) but is driven almost entirely by the misleading subset ($+11.40$); non-misleading Pass$\to$Fail cases actually show a decrease ($-3.64$) on average. This confirms that QA shortens successful repairs, while misleading QA can increase exploration effort before failure.

\subsection{Characterization of Generated QA}
\label{appendix:qa_characterization}

\begin{table*}[!t]
  \caption{Characterization of generated QA pairs across knowledge categories.}
  \centering
  \scriptsize
  \label{tab:qa_knowledge_summary}
  \setlength{\tabcolsep}{4pt}
  \renewcommand{\arraystretch}{0.98}
  \resizebox{\textwidth}{!}{%
  \begin{tabular}{@{} p{0.15\textwidth} c p{0.18\textwidth} p{0.25\textwidth} p{0.25\textwidth} @{}}
  \toprule
  \textbf{Category} & \textbf{Prop.} & \textbf{Prominent Knowledge} & \textbf{Typical Answer Signals} & \textbf{Repair Contribution} \\
  \midrule
  Mechanism \& Behavior &
  74.4\% &
  Logic flow, data transform, error \& exception &
  Explanation, code snippet, line reference, error \& trace &
  Finds causal fix points and clarifies behavior boundaries. \\
  \midrule
  Design \& Usage &
  13.9\% &
  API contract, design intent &
  Code entity, usage example, caveat &
  Constrains valid edits and preserves architectural consistency. \\
  \midrule
  Locating \& Structure &
  9.5\% &
  Implementation site, definition site &
  File path, code entity, line reference &
  Narrows search space and anchors dependencies. \\
  \midrule
  Ecosystem \& Standards &
  2.2\% &
  Protocol standard, library behavior &
  Caveat, usage example, summary &
  Prevents external inconsistency and regressions. \\
  \bottomrule
  \end{tabular}
  }
\vspace{-1.0em}
\end{table*}

Table~\ref{tab:qa_knowledge_summary} summarizes the knowledge carried by the generated QA pairs across four categories. The distribution is heavily skewed toward \emph{Mechanism \& Behavior} (74.4\%), reflecting that most issues demand understanding of internal logic flows, data transformations, and exception paths before a correct fix can be crafted. The remaining quarter is shared among \emph{Design \& Usage} (13.9\%), \emph{Locating \& Structure} (9.5\%), and \emph{Ecosystem \& Standards} (2.2\%), which together supply complementary knowledge dimensions such as API contracts, dependency anchoring, and external constraints.

Although the 44 Fail$\to$Pass instances follow a similar Mechanism-dominant distribution, 20 of the 44 instances (45.5\%) require at least one question from a non-Mechanism category: \emph{Design \& Usage} in 11 cases, \emph{Locating \& Structure} in 7, and \emph{Ecosystem \& Standards} in 2. This suggests that for a substantial portion of previously failed instances, Mechanism knowledge alone was insufficient, and cross-category combinations contributed to the successful fix, consistent with the coverage advantage reported in the category-guided ablation (Section~5.3 of the main paper).

\section{Human Audit and Regression Details}
\label{appendix:audit_regression}

This section provides supplementary details for the human audit and regression analysis reported in Section~5.2 of the main paper.

\subsection{Minor Deviation Breakdown}
\label{appendix:audit_details}

Among the 132 QA pairs labeled \emph{Supported with minor deviations}, we categorize the imperfections into three broad types:

\begin{itemize}[leftmargin=*,itemsep=2pt,topsep=2pt]
  \item \textbf{Localized reference or detail inaccuracies (67.9\%).}
    The main explanation is supported, but some local details such as
    code references, examples, function ownership, API details, or line
    ranges are imprecise.
  \item \textbf{Evidence-scope overreach (24.6\%).} The answer
    extrapolates beyond directly verifiable repository evidence, e.g.,
    by stating intent, broad causal explanations, external ecosystem
    behavior, or known limitations.
  \item \textbf{Implementation boundary overgeneralization (7.5\%).}
    The described mechanism is broadly correct, but the answer
    overgeneralizes special cases, branch-specific behavior, or
    boundary conditions.
\end{itemize}

\subsection{Cases Containing Ungrounded Claims}
\label{appendix:ungrounded_cases}

The two QA pairs labeled \emph{Contains ungrounded claim} still
contained useful repository-grounded information, but their central
mechanism claims were not supported by the code.

\begin{itemize}[leftmargin=*,itemsep=2pt,topsep=2pt]
  \item \texttt{django\_\_django-12663\_q0}. The answer correctly
    identified the relevant lookup-preparation path and field-preparation
    logic, but its central mechanism claim was wrong: it claimed that
    \texttt{SimpleLazyObject} inherits from \texttt{Promise} and would
    therefore be automatically evaluated by the \texttt{Promise} branch
    in \texttt{Field.get\_prep\_value()}. The repository instead defines
    \texttt{SimpleLazyObject} under the \texttt{LazyObject} hierarchy,
    not \texttt{Promise}, so this automatic \texttt{Promise}-based
    evaluation claim is contradicted by the code.
  \item \texttt{sympy\_\_sympy-23534\_q0}. The answer correctly located
    \texttt{symbols()} and described the normal use of \texttt{cls},
    but incorrectly claimed that \texttt{cls} is propagated through
    nested tuple recursion; the implementation recursively calls
    \texttt{symbols(name, **args)} without forwarding
    \texttt{cls=cls}.
\end{itemize}

\subsection{Pass-to-Fail Regression Details}
\label{appendix:p2f_details}

We manually inspect all 22 Pass$\to$Fail regression trajectories analyzed in Section~5.2 of the main paper to distinguish failures caused by misleading QA from failures where QA was not the primary cause of regression. We label a case as \emph{misleading} when the dominant downstream effect of the injected QA on the \resolvertag's repair behavior is misleading. Under this criterion, 5 of the 22 Pass$\to$Fail cases are misleading, while the remaining 17 are non-misleading.

\subsubsection{Misleading Cases}

We further categorize the 5 misleading Pass$\to$Fail cases by the \resolvertag's primary repair behavior, as summarized in Table~\ref{tab:p2f_misleading_behavior}.

\begin{table}[H]
  \caption{Repair-behavior breakdown of the 5 misleading Pass$\to$Fail cases under DeepSeek-V3.2.}
  \centering
  \small
  \setlength{\tabcolsep}{4pt}
  \begin{tabular}{@{}p{0.25\linewidth} c p{0.50\linewidth}@{}}
    \toprule
    \textbf{Repair behavior} & \textbf{\#} & \textbf{Relation to QA} \\
    \midrule
    Wrong modification location & 2 & QA emphasizes a related but
      non-gold location, which the \resolvertag keeps searching and
      editing. \\
    Wrong repair logic          & 2 & QA-provided mechanisms or context
      are turned into logic inconsistent with the gold fix. \\
    Overbroad edits             & 1 & QA provides an overly broad set
      of related methods or files, leading to unnecessary edits. \\
    \bottomrule
  \end{tabular}
  \label{tab:p2f_misleading_behavior}
\end{table}

In these cases, the QA often still contains repository-grounded facts, relevant code locations, or locally correct mechanism descriptions. The failure mode is that the QA provides a plausible but incorrect repair framing, and the \resolvertag repeatedly follows the QA-highlighted location, mechanism, or scope during search, editing, and testing, eventually deviating from the gold fix.

Cross-referencing with the answer-level audit (Section~\ref{appendix:audit_details}), among the 10 QA pairs in the 5 misleading cases, only 1 was labeled \emph{Contains ungrounded claim}; the remaining 9 were \emph{Supported}. The clearest ungrounded-claim example is \texttt{django\_\_django-12663\_q0} (see Section~\ref{appendix:ungrounded_cases}), where the wrong \texttt{SimpleLazyObject}/\texttt{Promise} inheritance claim makes the failure look like a lazy-object value-preparation issue and steers the \resolvertag toward the wrong repair direction. In contrast, many other misleading cases are more subtle: the QA may describe relevant repository mechanisms correctly, but those mechanisms are not the right repair target.

\subsubsection{Non-misleading Cases}

For the remaining 17 non-misleading Pass$\to$Fail cases, QA generally played a more constructive role: it often helped the \resolvertag identify relevant files, understand the failure mechanism, or make partial progress toward the gold fix. The eventual failures were more often due to \resolvertag-side behavior, such as overgeneralizing a correct clue, implementing only part of the required fix, adding extra non-gold behavior, or submitting a patch that did not preserve the QA-guided repair direction.

We note that the \resolvertag's prompt explicitly instructs it to treat injected QA as auxiliary context that must be cross-checked against the repository, and to discard or override QA whenever it conflicts with directly observed code, command outputs, or test feedback rather than blindly following it. Therefore, these cases should not be interpreted as QA misleading the \resolvertag; rather, they show limitations in how the \resolvertag uses otherwise helpful or partially helpful QA, even under a cautious-use prompting policy.

This regression analysis confirms a limitation of \approach: QA can influence repair behavior negatively in some cases. However, imperfect intermediate artifacts are an inherent limitation of all LLM-based methods, as pre-fix retrieval, plan-then-fix, agentic patch generation, and direct end-to-end repair are all subject to hallucinated or partially incorrect intermediate outputs, rather than a defect specific to QA injection. The audit and regression analysis quantify this risk concretely under \approach and show that misleading QA accounts for only 5 of the 22 inspected regressions, while the framework yields a net $+$22 instance gain, indicating that the risk is bounded and substantially outweighed by the benefit.

\end{document}